# STEADY-STATE FLOW-FORCE COMPENSATION IN A HYDRAULIC SPOOL VALVE

## Jan Lugowski

*Department of Mechanical Engineering Technology, Purdue University*
*401 N. Grant St., 47907-2021 West Lafayette, IN, USA, lugowskj@purdue.edu*

**Abstract**

A high-speed jet flowing inside of a partially-open hydraulic valve is accompanied by a reaction force, also referred to as flow force. The nature of this force has remained a mystery despite an extensive research effort spanning many decades. The momentum theory on the flow force by Lee and Blackburn (1952) explains the origin of the flow force and offers a design solution to shape the valve spool as a turbine bucket. It provides a model to calculate the compensated flow force as well. This paper shows that the model applies to a different flow case due to incorrect assumptions made. A corrected equation is presented based on a detailed analysis of the static-pressure distribution in the valve cavity as well as on a literature review of pressure loss in diffusers and nozzles. The new equation is based on the compensation taking place upstream of the valve orifice, not downstream as assumed by the momentum theory. The new model can be applied to chamfers or notches on the valve spool without the need to machine a complete turbine-bucket profile.

## 1 Introduction
### 1.1 Background and Hypothesis

In a partially-open hydraulic valve, shown in Fig. 1, a high speed jet flows into the spool cavity, generating an axial flow force $F_x$ that acts on the spool in the direction to close the valve. The spool cavity is shaped as a turbine bucket to reduce the magnitude of the flow force. A full, uncompensated, flow force $F_x$ exists in a square-land spool, Fig. 2a. The flow force $F$ is calculated from the momentum change of the fluid entering and leaving the control volume CV, as first proposed by Lee and Blackburn (1952), and reprinted by Blackburn et al. (1960), Merritt (1967) and Guillon (1969). The control volume includes the spool that can slide in the axial direction. The axial component of the flow force, $F_x$, acts on the spool to close the valve. The radial component $F_y$ acts on the valve body (sleeve) cancelling itself on the circumference, and also for that reason it does not affect the spool, see Fig. 2a. If the axial flow force $F_x$ is too large for direct operation of the spool, pilot-operated valves are required. Otherwise, the flow force needs to be reduced, or compensated. The above literature provides a design solution to this end by shaping the spool cavity in the form of a turbine bucket (Fig. 2b). Guillon (1969) considered the dynamic hydraulic forces an extremely difficult problem for which no satisfactory solution had been found, which is still true. Lugowski (1985, 1993) made measurements that the momentum theory could not explain; this paper is the continuation of that effort.

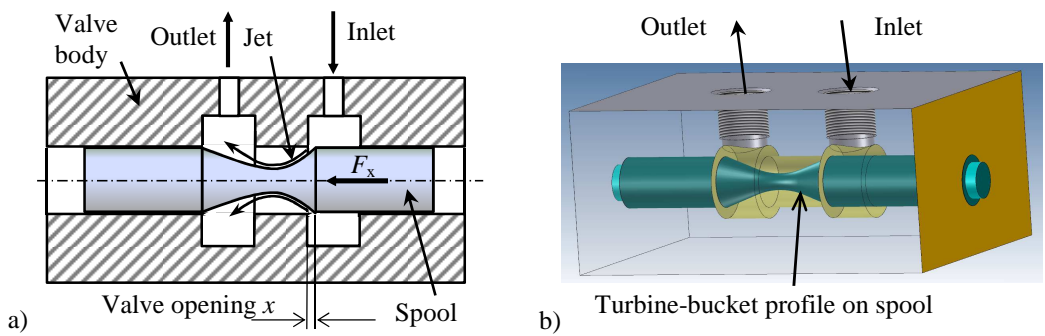

**Fig. 1:** *Cross-section (a) and isometric view (b) of a 2-way valve with a compensated spool*

The following hypothesis is the subject in this paper: At steady-state flow conditions, a fluid jet entering a spool cavity featuring a turbine-bucket profile loses its velocity in an area adjacent to the vena contracta, and thus is not capable of creating an opening, or compensating, flow force *downstream* from the vena contracta when the jet leaves the spool cavity. Instead, the compensation of the flow force takes place *upstream* from the vena contracta due to the unbalanced static pressure acting on the spool chamfer.

The momentum equation is analysed for flow-force compensation on a profiled spool in Part 2. The modified equation for flow-force compensation and a literature overview of diffusers as applied to hydraulic valves is presented in Part 3. The modified equation is experimentally verified with friction factors included to account for pressure losses due to friction in Part 4. The velocity field and pressure distribution in the valve orifice is presented in Part 5 using computational fluid dynamics (CFD) model of flow in the valve. The paper concludes with a summary.

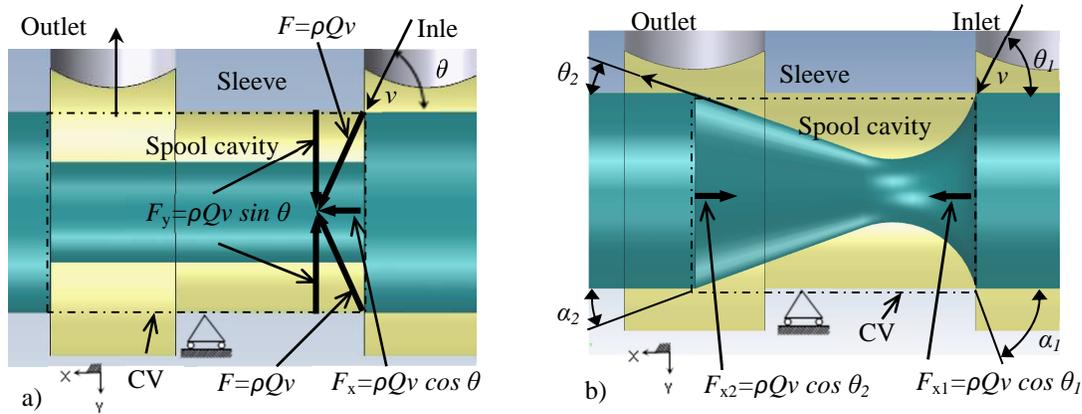

**Fig. 2:** *Cross-section of a valve with an uncompensated valve, where fluid enters the control volume (CV) at maximum velocity v and slows down to zero velocity inside of CV generating axial flow force $F_x$ acting on the spool to close the valve (a) and the compensated spool with a turbine-bucket profile along which the jet maintains its maximum velocity v, generating negative compensating flow force $F_{x2}$ as it leaves CV (b)*

## 1.2 Definitions

The Nomenclature listing in Part 5 defines all the symbols used. The conventions used in the paper are illustrated in Fig. 2. The coordinate system XY for the axial and radial forces indicates the *active* forces $F$ on the spool as per literature on the flow force. Staying with the convention, all the forces presented in this paper are depicted as active forces. Each corresponding reaction force $R$ would have a reversed direction (sign) and would keep the spool from moving, as per convention applied by Fox et al. (2011). The angle $\alpha$ refers to the slope of the spool profile while angle $\theta$ represents the jet angle, Fig. 2b. An axial flow force $F_x$ that acts in the direction to close the valve is considered positive. A force acting to open the valve is considered negative and is also referred to as the compensating force.

## 1.3 Related Work

Lugowski (1985) performed experiments to reduce the flow force on a spool with a turbine-bucket profile. The flow-force versus spool displacement matched that given by Lee and Blackburn (1952), but the compensation was not sufficient for the control system to work properly. Experiments with various angles of the turbine-bucket profile indicated that the exit angle $\alpha_2$, see Fig. 2b, did not contribute to the negative force generated by the profile. Even after the part of the spool containing angle $\alpha_2$ was cut-off, the profile featuring only entry angle $\alpha_1$ was still providing a similar flow-force versus displacement curve. To find the explanation for the effect of $\alpha_2$, the static-pressure distribution on angle $\alpha_1$ was recorded as presented by Lugowski (1993). This experiment showed that the compensation of the flow force occurs on chamfer $\alpha_1$, upstream from the vena contracta.

The accepted consensus has not changed since Guillon (1969) explained the flow force and its compensation. He made reference to the static pressure acting on the spool and contributing to the total axial flow force. He considered it a simple case that needed no further explanation, mentioning only the static pressure acting on the two end faces of the spool. In praxis, both spool ends are exposed to the same static pressure, such as a common drain or tank line. The effect of a high static pressure existing at the valve inlet port and acting on the entry profile $\alpha_1$ of the spool is considered here. The exit profile $\alpha_2$ of the spool is exposed to a lower valve-outlet pressure. The spool becomes statically unbalanced after the valve opens and when entry profile angle $\alpha_1 < 90°$, as in Fig. 2b. The resulting unbalanced static pressure on the spool profile is the source of the compensating flow force.

## 2   Where does the Compensation Take Place?

If only the dynamic jet forces are considered, as in Part 1.1, then the pressure distribution on the spool profile should be as shown in Fig. 3a. The pressure curve shows areas of pressure lower than $p_{out}$, where the positive flow force $F_{x1}$ and the negative, compensating, flow force $F_{x2}$ originate. As the jet enters the spool cavity, it also lowers the pressure on wall CE, which in turn produces a positive force $F_{x1}$ as the pressure on the opposite wall EG is higher. Jet leaving the spool cavity lowers the pressure on wall EG and produces the negative force $F_{x2}$. Two aspects related to the location of the origin of the flow forces are considered. The first aspect is that the axial flow force can be affected by an unbalanced static-pressure distribution on the spool profile outside of the control volume CV. Lugowski (1993) detected a significant source of the negative flow force $F_{x2}$ that acts to open the valve (Fig. 3b). The momentum theory, as shown in Fig. 3a, neglects forces from static pressure on wall BC because this wall lies outside of CV, see Fig. 2b. Since this wall is not cylindrical, pressure acting on it, when not counteracted by similar pressure on the opposite wall EG, generates a force. Even though the wall BC is very small, it is exposed to a very high valve inlet pressure and produces a negative force $F_{x2}$ (Fig. 3b) that can be equal to, or higher than, the positive flow force $F_{x1}$. The second, and related, aspect is that the compensating axial flow force $F_{x2}$ is generated *upstream* (Fig. 3b), not downstream (Fig. 3a) from the valve orifice.

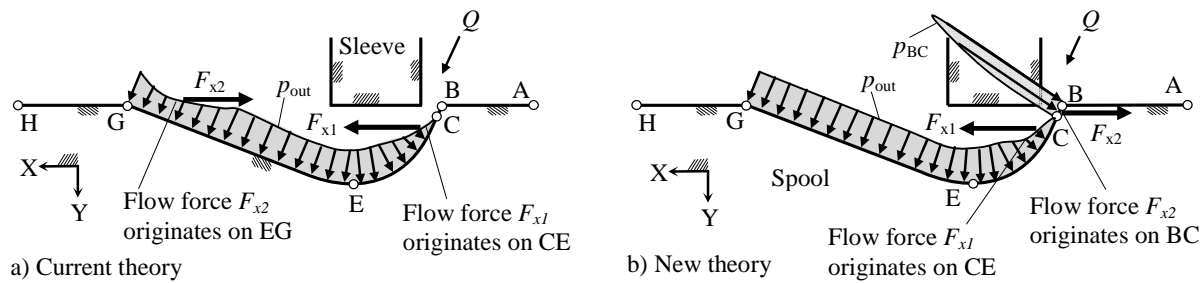

**Fig. 3:** *According to the momentum theory, a turbine-bucket profile produces the compensating flow force $F_{x2}$ on wall EG, (a), while the wall BC, which is exposed to high inlet pressure, has been omitted. Profile EG does not contribute to the flow-force compensation (b)*

When the valve is closed, the whole spool profile BCDEFG in Fig. 4a is under equal static pressure $p_{out}$. There is no net axial flow force, as the two forces due to pressure $p_{out}$ that pull the spool profile apart horizontally cancel out. If at any location on the spool profile BCDEFG the pressure differs from $p_{out}$, there will be a net axial flow force. If pressure is lower on wall BE, there will be a positive axial flow force that acts to close the valve because on the opposite wall EG the pressure is higher. Such is the case with a square-land spool shown in Fig. 4c. Both details (4b) and (4c) show the differential static pressure $p_{ds}$, obtained by subtracting pressure $p_{out}$ on wall EG from pressure on wall BD. The profile BD also depicts the pressure $p_{out}$. When $\alpha_1 < 90°$, Fig. 4b, the wall BC is under pressure that is higher than $p_{out}$. Pressure acting on wall BC acts to open the valve and contributes to the negative compensating flow force $F_{x2}$. Pressure on wall CD is lower than $p_{out}$ and contributes to the positive flow force $F_{x1}$. A square-land spool, shown in Fig. 4c, does not generate any pressure that is higher than $p_{out}$ on wall BD and thus does not have any compensation of the flow force.

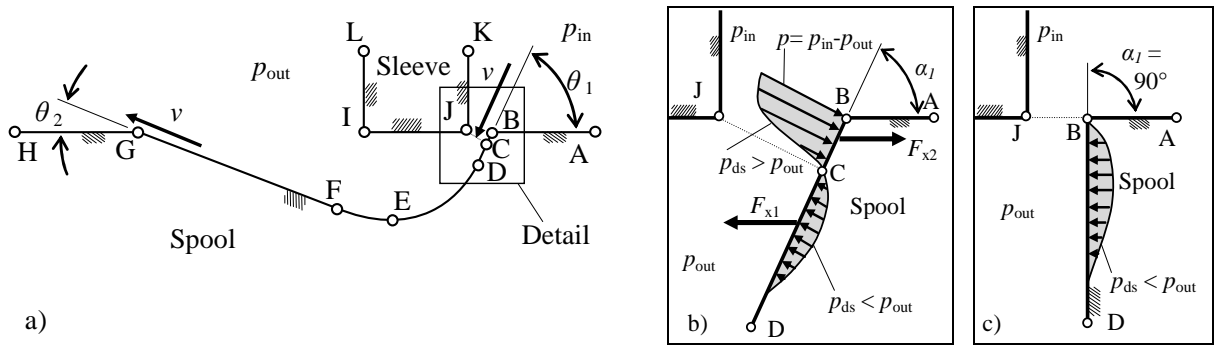

**Fig. 4:** *Per momentum theory, the compensation of the flow force on a turbine-bucket profile occurs as the jet exits at angle $\Theta_2$. Orifice area is shown enlarged in details b and c (a). The theory does not consider the pressure acting on profile BC ($\alpha_1 < 90°$) as contributing to the compensation (b); as is the case with an uncompensated spool ($\alpha_1 = 90°$) where orifice JC becomes JB (c)*

Such a model for force compensation has not been included in the momentum theory. As shown in Fig. 5, the theory assumes that the jet enters control volume CMNGIJC at angle $\theta_1$ and through orifice CJ, generates the closing force $F_{x1}$, follows the spool profile and leaves CV at angle $\theta_2$, producing the opening force $F_{x2}$. The magnitude of the resultant axial flow force $F_x$ is computed based on momentum influx and efflux as first proposed by Lee and Blackburn (1952):

$$F_x = F_{x1} - F_{x2} = \rho Q v \cos \Theta_1 - \rho Q v \cos \Theta_2 \qquad (1)$$

$$= \rho Q v (\cos \Theta_1 - \cos \Theta_2)$$

The positive axial component of the total flow force $F_x$ in Eq. (1) is a closing flow force, $F_{x1}$. The negative axial component is an opening, or compensating, flow force $F_{x2}$.

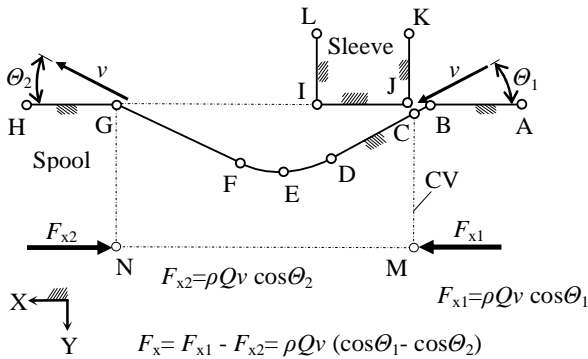

**Fig. 5:** *Axial flow forces acting on profiled spool due to momentum change of a jet entering and exiting the control volume*

Equation (1) assumes that the jet does not lose its velocity along the spool profile. The kinetic energy of the jet is maintained intact, as the jet velocity $v$ applies to both forces, $F_{x1}$ and $F_{x2}$. That means that the jet has a very high kinetic energy far away from the orifice at the valve outlet port, where the static pressure is usually much lower than at the valve inlet port. Lee and Blackburn (1952) considered Eq. (1) to give a maximum flow force under ideal conditions when there is no energy loss of the jet along the spool profile. If there were a velocity loss, then the negative flow force $F_{x2}$ would be smaller.

A spool with a turbine-bucket profile or with features like a chamfer or notch is not statically balanced when the valve is in an open position, as in Fig. 5. Another way to look at this configuration is to consider a spool with two different land diameters as in Fig. 6a. If diameter $D_1 < D_2$, then pressure $p_{out}$ will push the spool with a closing force $F_{x2}$ or act to close the valve. Depending on the level of pressure $p_{out}$ and the area of the annulus, the closing force can be significant, and such a valve would be statically unbalanced. The valve would close by itself if pressure $p_{out}$ was high enough to overcome the friction forces.

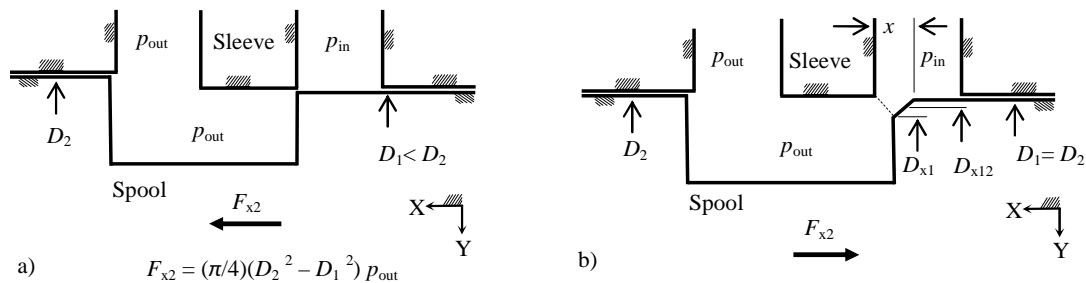

**Fig. 6:** *A statically-unbalanced spool with two different land diameters (a), and with a chamfered spool (b)*

A similar effect occurs on a chamfered spool (Fig. 6b), even though $D_1 = D_2$. If $p_{in} > p_{out}$ and the valve is open at $x$, then pressure $p_{in}$ acts not only on the cylindrical surface $D_1$, but also on the chamfer. As shown in Fig. 4b, pressure $p_{in}$ is not constant along the chamfer, but roughly drops from $p_{in}$ at $D_1$ to $p_{out}$ at $D_{x1}$, just above the vena contracta. The medium pressure acts on a diameter $D_{x12}$ that is located somewhere in-between. The force from the static imbalance $F_{x2}$ pushes the spool to open because pressure $p_{in}$ is higher than $p_{out}$.

## 2.1 Control Volume Assumptions

The momentum theory on the flow force in a hydraulic valve assumes a control volume as an entity on which forces act, without considering what happens inside of the control volume. This simplification may have led to incorrect conclusions regarding the negative compensating flow force $F_{x2}$. Three cases are considered to analyse this further. The first analysis is of flow forces by following the principles of fluid mechanics for a jet acting on a curved profile. What should happen inside of the control volume to satisfy Eq. (1) is considered next, concluding with the actual conditions existing there.

**Case 1:** Consider the control volume as a black box without regard to what happens to the jet inside of it. The momentum theory assumes that the jet maintains its velocity $v$ inside of control volume CV, see Fig. 5. The momentum change is due only to the change of the jet angle ($\Theta_1 + \Theta_2$), not due to the change of jet velocity $v$. By splitting the control volume CV in the middle at the lowest profile point E (Fig. 5), shown in Fig. 7, and analyzing the forces acting on the new control volumes, it follows that the positive flow force (closing the valve) is the force $F_{x2}$, see Fig. 7a. This force is larger when angle $\Theta_2$ is larger. However, Eq. (1) gives a larger positive flow force for the other angle, $\Theta_1$, and when $\Theta_1$ is smaller. For that reason Blackburn et al. (1960) suggested to make angle $\Theta_1$ large, equal to 69°, in order to achieve a smaller positive flow force on the profile. This positive flow force would be further compensated by making angle $\Theta_2$ small with a large negative flow force. The profile could over-compensate the flow force, so the resultant flow force $F_x$ would open the valve instead of closing it.

An uncompensated, square-land spool profile is shown in Fig. 2a where the axial flow force can be computed as follows:

$$F_x = \rho Q v \cos \Theta \tag{2}$$

The larger the angle $\Theta$, the smaller the flow force $F_x$ would be. Lee and Blackburn (1952) assumed this to be also true on a profiled spool, such as shown in Fig. 4a. There is, however, a difference in assumptions for a compensated spool and an uncompensated one. In the case of an uncompensated spool, the velocity $v$ of the jet entering the control volume is assumed to be reduced to zero in the axial direction inside of CV, or to be exiting the CV at 90°, see Fig. 2a and Eq. (2). No negative compensating flow force is present. For the compensated profile as shown in Fig. 4a, the authors assumed that the jet entering the control volume at velocity $v$ maintains it inside of the CV.

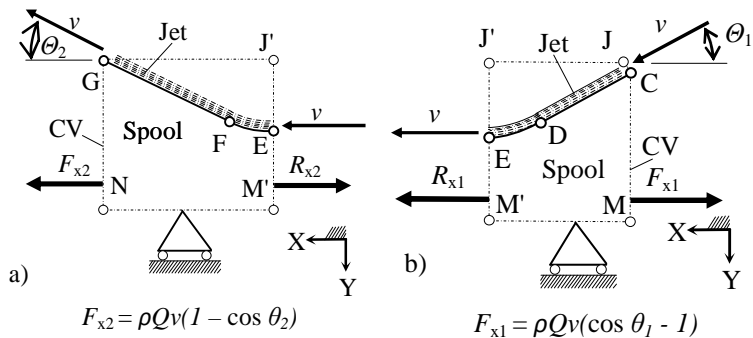

$F_{x2} = \rho Q v (1 - \cos \theta_2)$          $F_{x1} = \rho Q v (\cos \theta_1 - 1)$

**Fig. 7**: *Components of the axial flow force acting on the spool profile: a closing flow force $F_{x2}$ (a), and an opening (compensating) flow force $F_{x1}$ (b), have both a reversed direction, contrary to Eq. (1)*

This small difference in one assumption results in a significant change of the physical model describing how and where the flow forces are generated on a profiled spool. For an uncompensated spool profile, such as in Fig. 2a, a larger angle $\Theta$ results in a *smaller and positive* flow force $F_x$ that acts to close the valve. However, for a compensated spool profile, a larger angle $\Theta_1$ results in a *larger and negative* flow force $F_{x1}$ that acts to open the valve, see Fig. 7b, because the axial component of the entering momentum is smaller at larger $\Theta_1$, while the full axial force acts on CV when the jet leaves CV axially. Since the exiting momentum is larger than the one that enters CV, the resultant force $F_{x1}$ is negative and acts to open the valve. This does not agree with Eq. (1) and Fig. 5 where the flow force $F_{x1}$ is considered positive and acting to close the valve.

**Case 2:** Another interpretation of Eq. (1) could be that it describes a jet entering the control volume at velocity $v$ and angle $\Theta_1$, slowing down to (almost) zero velocity and generating the closing flow force $F_{x1}$ (Fig. 5). Then the jet accelerates back to velocity $v$ and exits control volume at angle $\Theta_2$, generating the compensating flow force $F_{x2}$. Such a case would require adding energy to accelerate the jet inside of control volume.

**Case 3:** Assume the velocity $v$ of the jet diminishes to (almost) zero along chamfer $\Theta_1$, and the slowed-down jet leaves the control volume at negligibly small velocity through a large cross-section, generating inside of control volume only one axial force that acts on the spool to close the valve:

$$F_{x1} = \rho Q v \cos \Theta_1 \tag{3}$$

The negative, or compensating, component of the momentum flow force cannot be generated within the control volume because the jet has (almost) no momentum left. The static-pressure distribution on the spool profile needs to be included to account for the compensating flow force.

## 3 A Statically-Unbalanced Spool

The compensating force $F_{x2}$ remains to be defined in a new way. Equation (1) does not account for this force, generated by the unbalanced static pressure acting on the spool profile BC, see Fig. 4b. Area BC is located outside of CV and is exposed to a high static pressure that is present at the valve inlet port. The static pressure $p_{BC}$ on the wall BC varies, see Fig. 4b. This pressure can be as high as the valve inlet pressure $p_{in}$ at the spool edge B and drop to the level even below the valve outlet pressure $p_{out}$ at C on the spool profile, as recorded by Lugowski (1993). The opposite wall EG of the profile is exposed to a lower, and constant, valve outlet pressure $p_{out}$ as discussed in Part 2.1. In the axial direction, the force $F_{x2}$ that results from pressure $p_{out}$ and $p_{BC}$ on the spool profile can be calculated from the differential static pressure $p_{ds}$ integrated over area $A$:

$$F_{x2} = -\int_A p_{ds} \, dA \tag{4}$$

The compensating flow force $F_{x2}$ as calculated from Eq. (4) replaces the dynamic compensating flow force $(-\rho Q v \cos \theta_2)$ in Eq. (1). As discussed in Part 2.1, the jet cannot generate this negative flow force on the spool profile downstream from the valve orifice. The negative sign means that $F_{x2}$ acts to open the valve, as it is the compensating flow force. The annulus area $A$, shown also as BC' in Fig. 8, is on a plane perpendicular to the spool axis. The differential static pressure $p_{ds}$ is defined as:

$$p_{ds} = p_{BC} - p_{out} \tag{5}$$

Pressure $p_{BC}$ pushes the spool horizontally to open the valve, while pressure $p_{out}$ on the opposite wall EG (Fig. 3b) pushes it to close the valve. The differential pressure $p_{ds}$ is positive because $p_{BC}$ is always higher than $p_{out}$. The static pressure $p_{BC}$ on the spool profile BC, if known, can be applied in Eq. (5). This pressure varies on wall BC as discussed above. If it is not available, it can be assumed that the average differential pressure $(p_{in} - p_{out})/2$ acts on the area BC', see Fig. 8. That area is proportional to the valve opening $x$. Based on this, Eq. (4) can be simplified as shown in Eq. (6). As the flow converges into the orifice, its velocity increases and, at higher flow rates Q, pressure $p_B$, calculated from Bernoulli's equation, at the spool control edge B will be somewhat lower than $p_{in}$. This pressure drop can be accounted for by substituting $p_B$ for $p_{in}$ in the formula for $F_{x2}$.

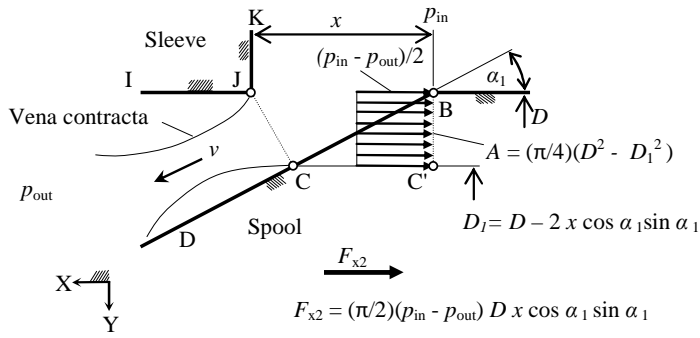

**Fig. 8:** *The compensating axial flow force $F_{x2}$ can be computed from the differential static-pressure distribution on the spool chamfer $\alpha_1$ upstream from the vena contracta*

$$F_{x2} = -\frac{\pi}{8}(p_{in} - p_{out})[D^2 - (D - 2x\cos\alpha_1 \sin\alpha_1)^2] \tag{6}$$

Since the term $2x\cos\alpha_1 \sin\alpha_1$ is usually very small compared to the spool diameter $D$, the above equation can be even more simplified:

$$F_{x2} = -\frac{\pi}{2}(p_{in} - p_{out}) D x \cos\alpha_1 \sin\alpha_1 \tag{7}$$

Further downstream from the valve orifice CJ, along the spool wall CD, the static pressure drops below the level of the valve outlet pressure $p_{out}$ as the jet accelerates further while it passes the vena contracta, as measured by Lugowski (1993). Since the pressure $p_{out}$ on the opposite spool wall EG (see Fig. 3b) is higher, a closing (positive) component of the axial flow force $F_{x1}$ is generated there. That dynamic force can be calculated from Eq. (1). However, the static flow force $F_{x2}$ is calculated as per Eq. (7). The total axial flow force $F$x on the spool can be calculated from Eq. (8), which replaces Eq. (1).

$$F_x = F_{x1} - F_{x2} = \rho Q v \cos\Theta_1 - \frac{\pi}{2}(p_{in} - p_{out}) D x \cos\alpha_1 \sin\alpha_1 \tag{8}$$

## 4  Friction Losses and Flow Force

The literature overview presented in the following supports the discussion in Part 2.1 which concluded that jet loses its velocity downstream from the valve orifice. The flow region immediately downstream from the vena contracta can be considered an asymmetric diffuser. Diffusers are used to recover the static pressure in the fluid by gradually increasing the cross-sectional area of the conduit, and thus decreasing the velocity of the fluid. Per Bernoulli's equation, the total energy of fluid at a given location is the sum of its kinetic energy (represented by dynamic pressure) and its potential energy or static pressure. The diffuser efficiency, or pressure recovery coefficient, is defined as the ratio of actual static pressure recovery and ideal static pressure recovery, that is when the whole kinetic energy at vena contracta converts fully to potential energy as static-pressure rise, see Fox et al. (2011). ASHRAE Handbook (1981) provides loss coefficient data for exit plane diffusers with angles 14° and above. The total pressure loss is smaller at smaller duct angles and larger exit/entry area ratios. For example, at area ratio 2, the loss is 37% at 14°, 50% at 30°, and 90% at 60°.

The literature on diffusers provides a velocity profile that is uniform across the whole section of a diffuser. There is no indication that an entry jet would flow along one wall of a diffuser without spreading. In order to achieve a reasonably good pressure recovery in a diffuser, a great design effort and an in-depth study of flow patterns are required. One can expect that in a hydraulic valve the high-energy jet having a very small cross-section area and flowing into an asymmetrical duct would separate from one wall, become irregular, and violently lose its energy, just as what happens in larger diffusers.

Spitzer (2001) describes the negative effect that valves, especially when partially closed, have on flow-rate measurements. Such valves create very complex flow patterns affecting all kinds of flow meters and should not be located upstream side of flow meters. Fox et al. (2011) describe losses in abrupt changes in area as primarily due to flow separation. In separated zones, energy is dissipated by violent mixing. A diffuser has typically turbulent flow and the static pressure rise in the direction of flow may cause flow separation from the walls if the diffuser is poorly designed.

Sovran (1967) provides detailed data for optimum geometries of diffusers, including diffusers with annular cross-section. In general, diffusers amplify non-uniformities of velocity-profile, while nozzles attenuate them. Large angles produce lower pressure recovery. The experimental data are for annular diffusers with dimensions of 15°-half angle, radius ratio from 0.55 to 0.70, and outer radius of the inlet annulus of 191.8 mm.

ASHRAE Handbook (2009) also indicates that at sharp entrances or sudden expansions separations of jet occur which produce large losses. A diffuser is used to reduce the loss in expansion but a separation may still occur, even a flow reversal (backflow), accompanied by excess losses. Streeter (1961) describes fully developed turbulent flow which changes greatly by comparatively small convergence or divergence of the walls. At diverging angles up to 6° self-preserving flows exist with local velocities that are inversely proportional to distance from the center line. For higher angles of divergence, flow separates from one wall and becomes asymmetric and irregular. Also, the presence of side walls tends to increase secondary flows which can cause flow separation. Such a flow becomes irregular and pressure recovery nearly ceases. Schneider et al. (2011) describe methods to enhance cross-sectional transport of high-momentum fluid in rectangular diffusers, and control of the location, shape, and size of the separation bubble. The numerical simulations they provide indicate a strong dependence of separation on the initial mean flow field.

Idelchik (1986) indicates that a transition from a larger section to a smaller one through a smoothly converging section (converging nozzles) is also accompanied by comparatively large irreversible losses of total pressure. Data he provided for rectilinear boundary walls, 90° angle of the converging section, and Reynolds number Re $\geq 10^5$, indicate the total pressure loss at about 0.2 of the maximum dynamic pressure at vena contracta. The optimal angle of the diverging section is 7°-10°, much smaller than found in hydraulic valves. To reduce pressure losses in the converging-diverging pieces, the converging nozzle should have an optimal bending radius of $R_{con}$= 0.5 – 1.0 $D_0$, where $D_0$ is the nozzle diameter. Diffuser losses were found to be: 0.48 at 30°, 0.65 at 45°, 0.76 at 60°, and 0.83 at 90° and above. Those losses are due to friction and turbulence, which means the kinetic energy of the jet is converted to heat.

Blackburn et al. (1960) discuss the losses in a valve in the context of flow instability and point out that a jet usually breaks up into a turbulent mass. Idelchik (1986) provides data for a discharge coefficient for a conical nozzle μ= 0.65-0.70, which is a measure of the jet contraction immediately downstream of the narrowest point of the nozzle. Further, the discussion of flow through an entry nozzle provides a formula to calculate the distance at which the flow becomes turbulent, and mentions the laminar boundary layer forming at the walls of the nozzle. At angles above 10°, the flow separates from the walls after passing the contracting section, and the separation is the main source of local losses of total pressure. In conclusion, the jet gets separated from the nozzle walls (see Fig. 8) and reattaches itself further downstream. As discussed earlier, the reattachment may be happening only on one wall.

Equation (8) needs to be modified to account for the pressure loss, as in diffusers and nozzles discussed above, by adding friction factors $f_k$, see Eq. (9). Compared to Eq. (8), the actual pressure acting on the spool profile is lower and generates smaller flow forces $F_{x1}$ and $F_{x2}$ as a result of friction losses in the jet. Due to the shortage of published data on pressure distribution in hydraulic valves, an attempt has been made to calculate the friction factors back to flow forces $F_{x1}$ and $F_{x2}$ resulting from experimental data given in Table 1, as published by Lugowski (1993). The friction factor $f_{k1}$ was found to be 0.367 and 0.289 at valve opening 0.08 mm and 0.23 mm, respectively, see Table 1. Considering that the flow forces $F_{x1}$ and $F_{x2}$ given in Table 1 were calculated, not measured, based on a total of eight pressure data points (shown plotted in Fig. 11), the variation in $f_{k1}$ was small at 0.078. This was not the case with $f_{k2}$ which was 0.250 and 0.724 at the valve opening 0.08 mm and 0.23 mm, respectively. At $x$= 0.08 mm the compensating flow force $F_{x2}$ was much smaller than the closing flow force $F_{x1}$. At the larger valve opening, $x$= 0.23 mm, the compensating force $F_{x2}$ was slightly larger, so that the total flow force $F_x$ became negative and acted to open the valve. Additional experimental data are needed to more precisely define the friction factors $f_{k1}$ and $f_{k2}$.

$$F_x = F_{x1} - F_{x2} = f_{k1}\, \rho Q v \cos \Theta_1 \qquad (9)$$

$$- f_{k2}\, \frac{\pi}{2}(p_{in} - p_{out})\, D\, x \cos \alpha_1 \sin \alpha_1$$

**Table 1:** Data from paper by Lugowski (1993) used for the experimental verification of Eq. (9)

| Parameter | x= 0.08 mm | x= 0.23 mm |
|---|---|---|
| $\rho$, kg/m$^3$ | 870 | 870 |
| $Q$, m$^3$/s | $1.200*10^{-4}$ | $1.667*10^{-4}$ |
| $v$, m/s | 39.0 | 18.8 |
| $\alpha_1$, deg. | 26 | 26 |
| $\Theta_1$, deg. | 26 | 26 |
| $D$, mm | 28 | 28 |
| $p_{in}$, kPa | 898 | 337 |
| $p_{out}$, kPa | 0 | 0 |
| $f_{k1}$ | 0.367 | 0.289 |
| $f_{k2}$ | 0.250 | 0.724 |
| $F_{x1}$, N | 1.34 | 0.71 |
| $F_{x2}$, N | 0.31 | 0.97 |
| $F_x$, N | 1.03 | -0.26 |

## 5  CFD Model

The parameters used by Lugowski (1993), as shown in Table 1, were also used to build the finite-element model of the valve by using multiphysics software COMSOL 4.3. A fluid-structure interaction model was based on the geometry of the orifice area of the test valve. The model was used to solve for the velocity field of the fluid in the orifice and for the fluid velocity and pressure on the spool profile. A 2-D model was used as the flow in one axial cross-section of the valve orifice is practically identical to the flow in any other axial cross-section. The inlet and outlet ports of the valve are large and far away compared with the size of the valve orifice so that their location and size provide the uniform flow through the orifice. As most of the pressure drop occurs in the valve orifice area, the selection of the model is appropriate.

The geometry of the whole model and the mesh in the orifice area are shown in Fig. 9. At the inlet of the model, fluid velocity was set to match the flow rate given in Table 1. At the outlet of the model the pressure was set to zero (atmospheric pressure), same as in Table 1.

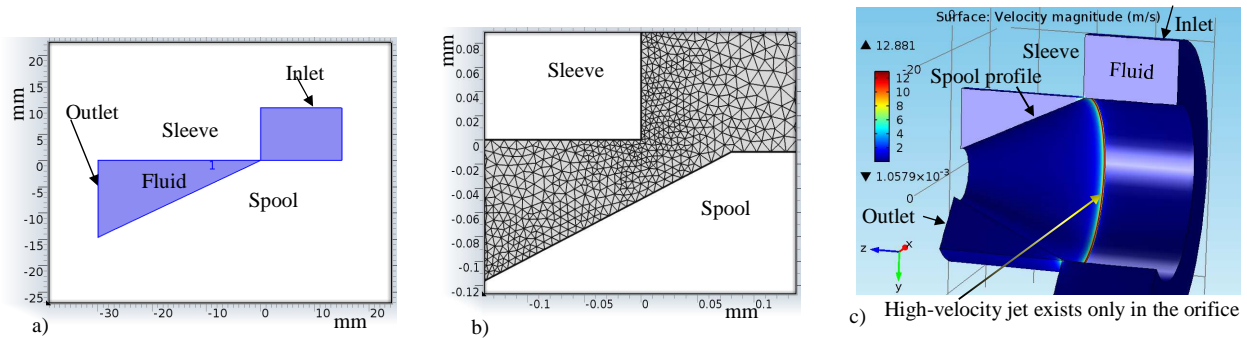

**Fig. 9:** *2-D geometry used for finite element model in COMSOL (a), the free triangular mesh shown for orifice area (b), and the velocity magnitude field for the 2-D axisymmetric geometry used for the comparison of computations (c)*

The calculated velocity field in the flow through the valve orifice at two valve openings, x= 0.08 mm, and x= 0.23 mm are shown in Fig. 10. Two lines were added to indicate the location of the spool control edge (point B) and sleeve (valve body) control edge (point J). The velocity of the fluid is shown to sharply decrease right downstream from the orifice, and then to spread out uniformly across the cross-section between the sleeve and spool. The arrow-surface graphs show the spread of the fluid even more clearly. There was no sign of a jet maintaining its maximum velocity $v$ along spool profile as assumed by Eq. (1). Since the orifice area is very small compared with the size of the flow model, the velocity field in the orifice area was visible only as a small red dot on a blue background. If the jet maintained its maximum velocity, there would be a red line along the spool profile. A red line of high jet velocity is only visible in Fig. 9c, where this line depicts the orifice along the spool circumference. Also in this model, there was no sign of a high-velocity jet along the spool profile.

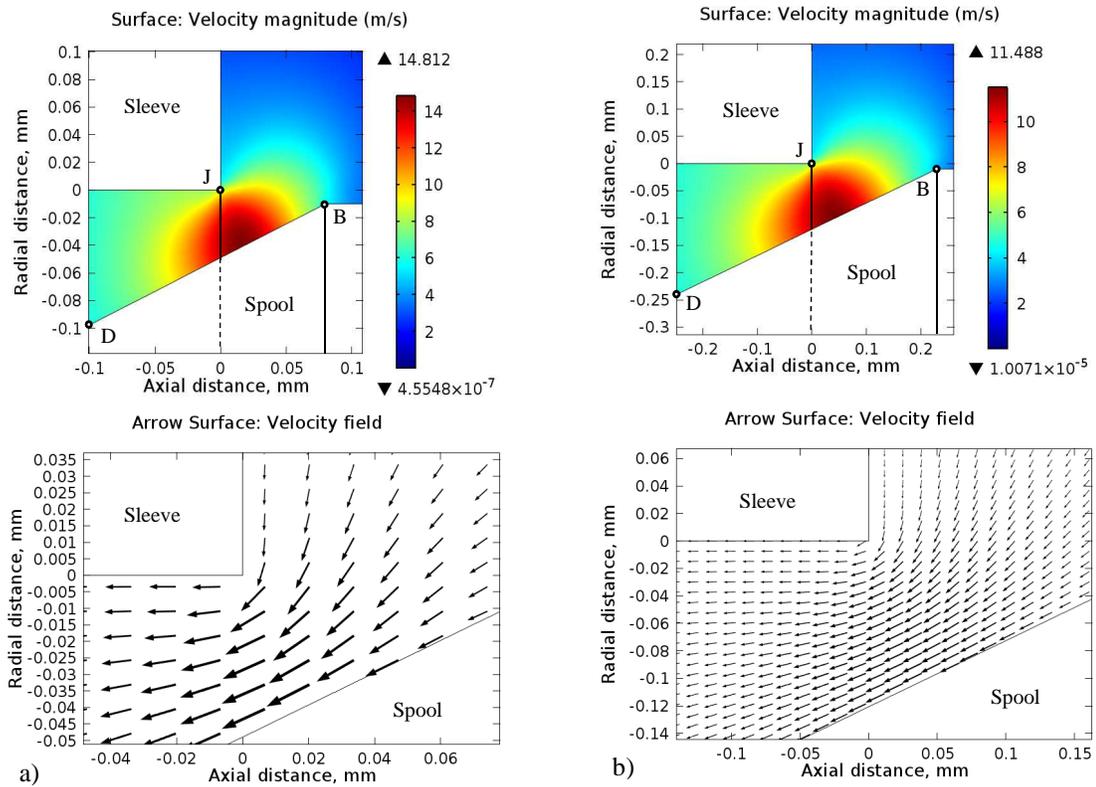

**Fig. 10:** *Velocity field of fluid in the valve orifice with parameters per Table 1: valve open at x= 0.08 mm (a) and at x= 0.23 mm (b)*

The calculated fluid velocity and pressure on the spool profile for the 2-D model, which produced similar results as the 2-D axisymmetric model, are shown in Fig. 11. A cutline was used in COMSOL to plot the flow parameters along profile BD of the spool. The straight cutline was extended by 0.4 mm beyond point B on the spool to show the velocity and pressure upstream from the orifice. Two edges for sleeve and spool were superimposed on the graphs to help locate the valve orifice. The velocity graphs show a sharp decrease of fluid velocity downstream from the valve orifice. This result further confirms that the jet loses its kinetic energy due to turbulence and friction in the area immediately downstream from the orifice. Furthermore, there was no indication of any pressure recovery in the diffuser region. Over there, fluid pressure acting on the spool profile became negative over a small area and then increased only to zero without a further recovery. Positive pressure between edges B and J, where the jet accelerates, produces the negative compensating flow force $F_{x2}$ acting on the spool. Negative pressure on the spool profile results in the positive flow force $F_{x1}$ as the jet decelerates further downstream of the valve orifice.

The pressure distribution as measured by Lugowski (1993) is shown in Fig. 11 to compare with the COMSOL calculations. At the larger valve opening, Fig. 11d, there is a relatively good agreement between calculation and experiment. Both the experimental data and the CFD model support the hypothesis that the jet loses its velocity immediately downstream from the orifice. This means that the modified equation for the flow-force compensation presented above is closer to represent the physics of the fluid flow in a hydraulic valve than the current equation.

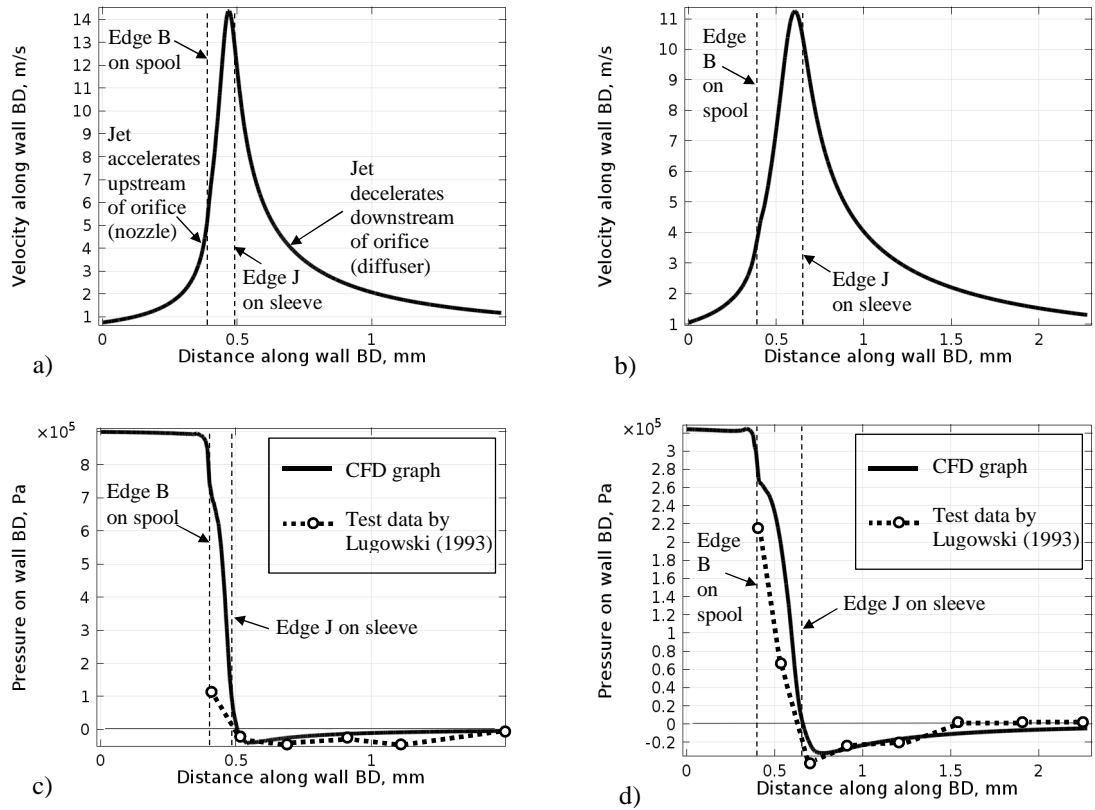

**Fig. 11:** *Final-element calculations for the velocity of fluid and the pressure along spool wall BD, see also Figs. 4 and 10, with the cutline in COMSOL model extended by 0.4 mm beyond the spool control edge (point B); All corresponding parameters as in Fig. 10: Fluid velocity at x=0.08 mm (a), and at x=0.23 mm (b), pressure on wall BD of spool at x= 0.08 mm (c), and at x= 0.23 mm (d)*

## 6 Summary

A spool featuring a turbine-bucket profile becomes statically unbalanced when a part of chamfer $\alpha_1$ (Fig. 4b) is exposed to a high valve-inlet pressure. This results in an opening (compensating) force. The static imbalance of the spool profile was not included by the momentum theory on the origin of the axial flow forces acting on the spool. A mathematical model is presented to account for this force.

A review of the literature on diffusers indicates that the jet loses its velocity significantly after passing the orifice due to turbulence, wall detachment and mixing with the surrounding fluid. CFD calculations and experimental data also support this approach. Total pressure loss in the jet starts already upstream of the orifice. As the pressure losses affect the static-pressure distribution on the spool profile and the flow forces, they have been accounted for by introducing the friction factors $f_k$.

**Nomenclature**

| | |
|---|---|
| $\alpha_1$ | Entry profile (chamfer) angle, deg. |
| $\alpha_2$ | Exit profile (chamfer) angle, deg. |
| $\rho$ | Mass density of fluid, kg m$^{-3}$ |
| $\Theta$, | Jet angle, deg. |
| $\Theta_1$ | Entry jet angle, deg. |
| $\Theta_2$ | Exit jet angle, deg. |
| $A$ | Spool profile area (projected on a plane perpendicular to spool axis), m$^2$ |
| $D$ | Spool diameter, m |
| $f_k$ | Dimensionless friction factor (loss of kinetic energy) |
| $f_{k1}$ | Friction factor for positive flow force |
| $f_{k2}$ | Friction factor for negative (compensating) flow force |
| $F$ | Flow force, N |
| $F_x$ | Axial flow force, N |
| $F_{x1}$ | Axial flow force to close the valve (positive), N |
| $F_{x2}$ | Axial flow force to open the valve (negative, compensating), N |
| $F_y$ | Radial flow force, N |
| $p_{ds}$ | Differential static pressure, Pa |
| $p_{in}$ | Valve inlet pressure, Pa |
| $p_{out}$ | Valve outlet pressure, Pa |
| $p_s$ | Static pressure, Pa |
| $Q$ | Flow rate, m$^3$s$^{-1}$ |
| $R$ | Reaction flow force, N |
| $Re$ | Reynolds number |
| $v$ | Velocity of jet at vena contracta, m s$^{-1}$ |
| $x$ | Valve opening (axial), m |

**Acknowledgements**

This work was sponsored by Purdue University's Maha Fluid Power Educational Advisory Board and by Parker Hannifin Corporation, Hydraulic Valve Division. The author is grateful to his sons Adam Lugowski and Jan (John) Lugowski, and colleagues Nancy Denton and Gozdem Kilaz from Purdue University for their encouragement and revisions to the manuscript.